\documentclass{iau} 
\usepackage{graphicx}

\title[Binary CS of Hf 38] 
{The Binary Central Star of Hf 38}

\author[Helen Barker] 
{Helen Barker$^1$} 

\affiliation{$^1$Dept. of Astronomy \& Astrophysics, The University of Manchester \\ Oxford Road, Manchester, UK \\
email: {\tt helen.barker@postgrad.manchester.ac.uk} \\[\affilskip]}

\pubyear{2017}
\volume{xxx}  
\jname{Title of your IAU Symposium}
\editors{A.C. Editor, B.D. Editor \& C.E. Editor, eds.}
\begin{document}

\maketitle

\begin{abstract}

Despite years of effort, the impact of central star binarity on planetary nebula formation and shaping remains unclear. This is hampered by the fact that detecting central star binarity is inherently difficult, and requires very precise observations. The fraction of planetary nebulae with binary central stars therefore remains elusive. 
This work presents initial results of central star analysis using data from the  VST H$\alpha$ Survey of the Southern Galactic Plane and Bulge (VPHAS+). The true central star of PN Hf 38 has been revealed, and it exhibits a 0.465$\pm$0.334 $i$ band magnitude excess, indicative of a M0V companion.

\end{abstract}

\section{Introduction}

Some planetary nebulae (PNe) are observed to have binary central stars (CS), and there is increasing evidence that the presence of a binary companion can determine the PN morphology (\cite{hillwig2016}). However, the total fraction of PN with binary CS remains unclear. This is due to the intrinsic difficulty in detecting binarity, even when high quality data is available. 

One method of detecting CS binarity is to search for a percent-level CS $i$ band excess, indicative of a low mass main sequence binary companion (\cite{demarco2013}). VPHAS+ (\cite{drew2014}) provides sub-arcsecond aperture photometry down to ~20th magnitude in the Sloan $u$, $g$, $r$ and $i$ bands, and a H$\alpha$ filter. This is a similar limiting brightness to the SuperCOSMOS H$\alpha$ Survey (\cite{parker2005}) from which the Macquarie/AAO/Strasbourg H$\alpha$ (MASH) catalogue (\cite{parker2006}, \cite{miszalski2008}) was created, however VPHAS+ should provide superior resolution. Once complete, there will be 1427 currently known PN within the VPHAS+ footprint. Presented here are the first results from analysis of the currently available VPHAS+ data (release 2).

\section{The Central Star of Hf 38}

\begin{figure}[b]
\begin{center}
\includegraphics[width=\linewidth]{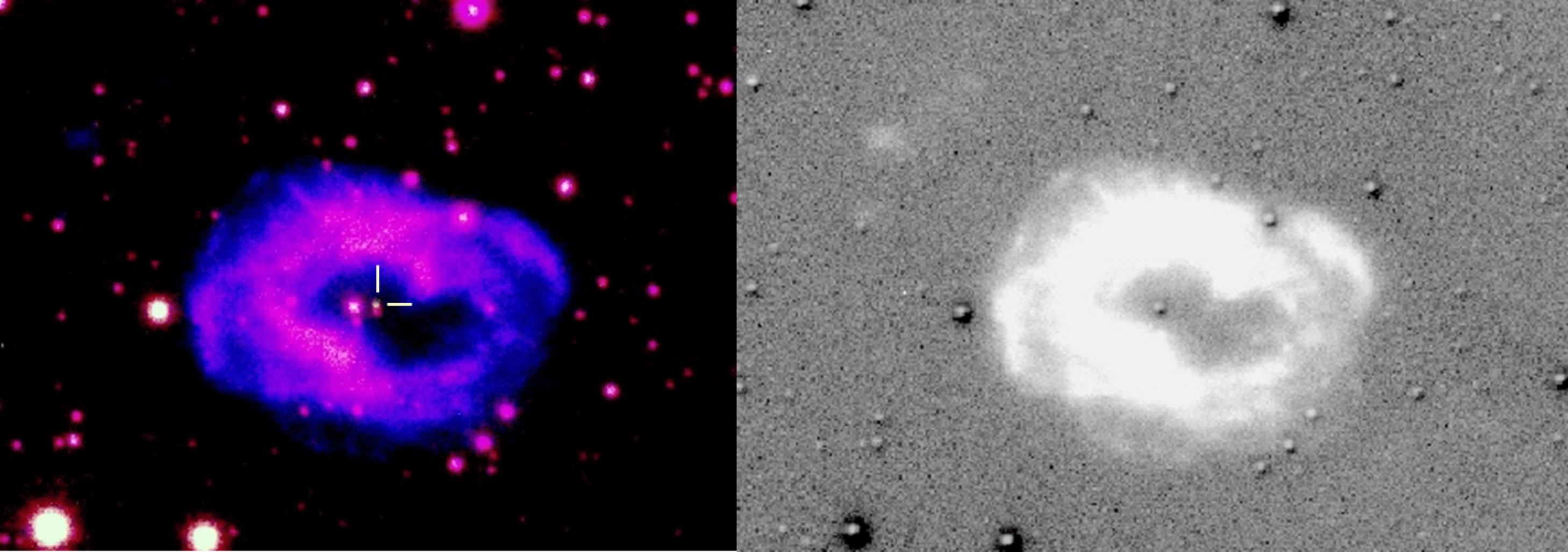} 
\caption{Images of Hf 38 from VPHAS+. On the left, $r$ band, $u$ band and H$\alpha$ data are shown in red, green and blue respectively  The right image show the H$\alpha$/$r$ quotient image.}
\label{hf38}
\end{center}
\end{figure}

The new, high resolution images of Hf 38 from VPHAS+ have revealed a faint blue star in the centre of the nebula. This is marked in the left panel of Figure \ref{hf38}. The right panel shows the H$\alpha$/$r$ quotient image, which highlights an arc of nebulosity ~0.36pc north west of the nebula, assuming a distance of 2.25kpc (\cite{frew2016}).

The marked star was confirmed to be the true CS by comparing the observed and expected colours. The expected CS colours were calculated by convolving the spectrum of a 100kK, log(g)=7.0, solar metallicity synthetic post-AGB atmosphere from TMAP (\cite{werner2003}) with the VPHAS+ bandpasses. The synthetic and observed colours were consistent within the errors. 

The reddening of this PN was determined by calculating E(B-V) using three methods. First, by comparing the observed to the synthetic $u-g$ colour. Second, by calculating the H$\alpha$/H$\beta$ line ratio using the spectrum from \cite{acker1992}. Third, by calculating the flux ratio using literature values of H$\beta$ (\cite{acker1992}) and radio flux (\cite{murphy2007}). These calculated values were compared to those in the literature. The calculated and literature E(B-V) values averaged to 0.90$\pm$0.08.

The observed CS colours were de-reddened using this value. The synthetic CS spectrum was scaled so the synthetic $g$ band magnitude matched the de-reddened one. The total flux of this spectrum was then calculated. At a distance of 2.25kpc, this resulted in a CS luminosity estimate of 1070L$_\odot$, consistent with the expected value.

The de-reddened $g-i$ colour was then compared to the synthetic ones. While the $g$ band will be unaffected by a low-mass main sequence companion, the companion's spectrum would peak in the $i$ band, thereby increasing the observed $g-i$ colour. An $i$-band excess of 0.465$\pm$0.334 magnitudes was detected in the CS of Hf 38, consistent with a M0V binary companion.



\begin{thebibliography}{}
	
\bibitem[Acker \etal\ 1992]{acker1992}
{Acker, A., Marcout, J., Ochsenbein, F., Stenholm, B., Tylenda, R., Schohn, C.}, 1992, \textit{The Strasbourg-ESO Catalogue of Galactic Planetary Nebulae. Parts I, II}	

\bibitem[Amari \etal\ 1995]{Amari_etal95}
{Amari, S., Hoppe, P., Zinner, E., \& Lewis R.S.}m 1995,
\textit{Meteoritics}, 30, 490 

\bibitem[De Marco \etal\ 2013]{demarco2013}
{De Marco, O., Passy, J.-C., Frew, D.~J., Moe, M., Jacoby, G.~H.}, 2013,
\textit{MNRAS}, 428, 2118 


\bibitem[Drew \etal\ 2014]{drew2014}
{Drew, J. ~E., Gonzalez-Solares,E., Greimel, R., Irwin, M., ~J., K{\"u}pc{\"u} Yoldas, A, Lewis, J., Barentsen, G., Eisl{\"o}ffel, J., Farnhill, H., ~J. }, 2014, \textit{MNRAS}, 440, 2036

\bibitem[Frew \etal\ 2016]{frew2016}
{Frew, D., ~J., Parker, Q, ~A., Boji{\v c}i{\'c}, I.~S.}, 2016, \textit{MNRAS}, 455, 1459

\bibitem[Hillwig \etal\ 2016]{hillwig2016}
{Hillwig, T., Jones, D., De Marco, O., Bond, H., Margheim, S., Frew, D.}, 2016, \textit{ArXiv e-prints}: {1609.02185}


\bibitem[Miszalski \etal\ 2008]{miszalski2008}
{Miszalski, B., Parker, Q. ~A., Acker, A., Birkby, J. ~L., Frew, D. ~J., Kovacevic, A.}, 2008, \textit{MNRAS}, 384, 525

\bibitem[Murphy \etal\ 2007]{murphy2007}
{Murphy, T., Mauch, T., Green, A., Hunstead, R.~W., Piestrzynska, B., Kels, A.~P., Sztajer, P.}, 2007, \textit{MNRAS}, 382, 382 

\bibitem[Parker \etal\ 2005]{parker2005}
{Parker, Q.~A., Phillipps, S., Pierce, M. ~J., Hartley, M., Hambly, N. C., Read, M. ~A., MacGilliveray, H. ~T.}. 2005, \textit{MNRAS}, 362, 689 

\bibitem[Parker \etal\ 2006]{parker2006}
{Parker, Q.~A., Acker, A., Frew, D.~J., Hartley, M., Peyaud, A.~E.~J., Ochsenbein, F., Phillipps, S., Russeil, D., Beaulieu, S.~F., Cohen, M., K{\"o}ppen, J.,  Miszalski, B., Morgan, D.~H., Morris, R.~A.~H., Pierce, M.~J., Vaughan, A.~E.}, 2006, \textit{MNRAS}, 373, 79



\bibitem[Werner \etal\ 2003]{werner2003}
{Werner, K., Deetjen, J.~L., Dreizler, S., Nagel T., Rauch, T., Schuh, S.~L.}, 2003, \textit{Stellar Atmosphere Modeling}, 288, 31


\end{thebibliography}
\end{document}